# BOTSPOT: DEEP LEARNING CLASSIFICATION OF BOT ACCOUNTS WITHIN TWITTER

*Christopher Braker[1], Stavros Shiaeles[2], Gueltoum Bendiab[2], Nick Savage[2], and Konstantinos Limniotis[3]*

[1] *CSCAN, University of Plymouth, PL4 8AA, Plymouth, UK*
*chris.brake@plymouth.ac.uk*

[2] *Cyber Security Research Group, University of Portsmouth, PO1 2UP, Portsmouth, UK*
*gueltoum.bendiab@port.ac.uk, sshiaeles@ieee.org, nick.savage@port.ac.uk*

[3] *National and Kapodistrian University of Athens, Greece*
*klimn@di.uoa.gr*

**ABSTRACT**. The openness feature of Twitter allows programs to generate and control Twitter accounts automatically via the Twitter API. These accounts, which are known as "bots", can automatically perform actions such as tweeting, re-tweeting, following, unfollowing, or direct messaging other accounts, just like real people. They can also conduct malicious tasks such as spreading of fake news, spams, malicious soft- ware and other cyber-crimes. In this paper, we introduce a novel bot detection approach using deep learning, with the Multi-layer Perceptron Neural Networks and nine features of a bot account. A web crawler is developed to automatically collect data from public Twitter accounts and build the testing and training datasets, with 860 samples of human and bot accounts. After the initial training is done, the Multi-layer Perceptron Neural Networks achieved an overall accuracy rate of 92%, which proves the performance of the proposed approach.

## I. INTRODUCTION

Twitter is currently one of the most used social media networks with over 326 million of monthly active registered users in 2019, where 46% of them are on the platform daily, together sending over 6,000 tweets every second, which corresponds to over 350,000 tweets per minute and 500 million tweets per day [1]. This social network provides a powerful micro-blogging platform where people can easily share their thoughts, feelings, and opinions about a wide variety of topics in the form of tweets (a message or a post limited to 140 characters) [9, 6]. Users can also share tweets created by other persons on their own Twitter feeds, which will be then visible to their friends and followers [13]. One other important feature of Twitter is the use of the hashtag, which is a word, or a phrase preceded by the pound sign (#) [14].

Hashtags help users to arrange and sort their tweets and discover other Twitter users who are interested in the same subjects [14]. With this feature, Twitter creates an online social structure consisting of clusters of interconnecting people, with celebrities having huge numbers of followers [11]. The potential capability of this platform in exchanging information in the form of views, thoughts, and opinions, makes it an ideal place for promoting a variety of relatively harmless tasks [9] such as targeted marketing or spreading fake news for malicious intent and manipulate the public opinion [10]. These activities are usually carried out by using fake accounts known as "bots" or "Sybils".

A bot is an automated user account controlled by a computer program that has been coded to interact with Twitters' API to perform automated tasks such as tweeting, following, unfollowing, or direct messaging other accounts. More sophisticated bots can even interact with human users, just like real people [6]. A recent study by the University of Southern California and Indiana University [12], found that up to 17% of Twitter active accounts are in fact bots rather than human users, which correspond to nearly 48 million accounts. Another study by Gartner [5] estimates that by 2020, 85% of customer requests will be handled by bots, while Inbenta estimates 1.8 billion unique customer chatbot users by 2021 [5]. Although some bot accounts such as Earthquake Bot are beneficial to the community, many are malicious. A study by Jordan Wright et al [18] confirmed that generated bot accounts are mainly used to spread spam and malware as well as influencing online discussion and sentiment. For example, bots have been used to redirect users to phishing websites [5], sway political elections, manipulate the stock market and create fake followers to make some people appear more popular than they are.

In this paper, we aim to address this issue by introducing a novel approach that uses Deep Learning to detect bot accounts. The proposed approach relies on the Multi-layer Perceptron Neural Network and a set of nine features that include account, tweets, and graph-related features. The Neural Network is trained on a dataset that consists of 760 normal and bot twitter accounts. Similarly, it was tested on a dataset of 100 twitter accounts either bot or normal. The initial experiments show promising results with 92% accuracy. The remainder of this paper is structured as follows. Section II gives an overview of existing Twitter bots detection techniques, their advantages, and drawbacks. In Section III, we present the methodology of the proposed method. Section VI presents the results of the experiments. Finally, Section V concludes the paper and presents future work.

## II. RELATED WORKS

The detection of bot accounts within Twitter has been an area of research that has gained more interest over the past few years and several approaches have been proposed to detect bots on Twitter. A recent review by Emilio Ferrara [6] classified those methods in three main categories; (a) methods based on social networks graphs, (b) systems based on crowd-sourcing and human computation, and (c) machine learning algorithms based on features of bots. In this section, we focus on the last category because our work falls in this category.

Machine learning-based approaches focus on the behavioural patterns of bot accounts that can be encoded in features and adopted with machine learning techniques to distinguish a bot account from a human one. For instance, Lin and Huang [9] proposed a method to detect bots in Twitter based on two Twitter account-related features: URL rate and interaction rate. The URL rate designs the number of tweets with URL in the total number of tweets, while interaction rate defines the ratio of the number of tweets interacting over the total number of tweets. The proposed approach was evaluated on a dataset that was collected from 26,758 public accounts with 508,403 tweets. Authors reported a precision rate between 82.9% and 88.5%, with the classification algorithm J48.

In [15], Song et al built a graph model to represent Twitter users and their relation-ships, where users and tweets represent the nodes of the graph and relationships represent the links between nodes. Then, distance and connectivity features are extracted from the graph and used to detect bot accounts. The distance defines the length of the shortest path between the tweet's sender and mentions, while the connectivity feature defines the strength of the connection between users. Authors claim that unlike account features, relation or graph-based features are difficult for spammers to manipulate and can be collected immediately. This approach achieved nearly 92% true positive with Bagging, LibSVM, FT, J48 and BayesNet classifiers. In more recent work [17], Alex Hai Wang used three graph-based features and three tweets-based features to facilitate spambots detection. The graph-based features (i.e., the number of friends, the number of followers and follower ratio) are extracted from the user social graph, while the tweets-based features (i.e., the number of duplicate tweets, the number of HTTP links, and the number of replies/mentions) are extracted from user's most recent 20 tweets.

The dataset used to evaluate this approach contains 25,847 users, around 500K tweets, and around 49M follower/friend that are collected from publicly avail-able data on Twitter. Different classification methods are used to identify spam bots including Decision Tree (DT), Neural Network (NN), Support Vector Ma-chines (SVM), Naive Bayesian (NB) and K-Nearest Neighbours (KNN). The NB classifier achieved the best results with 91% accuracy, 91% recall and 91% F-measure. In the same context [13], Lacopo Pozzana et al studied the behavioural dynamics that bots exhibit for one activity session via four tweet features: mentions per tweet, the text tweet length, fraction of retweets and fraction of replies. This study found behavioural differences between human users and bot account, which can be exploited to improve bot detection techniques. For example, over the course of their online activity, humans are constantly exposed to posts and messages by other users, so their probability to engage in social interaction in- creases. Authors used five machine-learning algorithms (DT, Extra Trees (ET), Random Forests (RF), Adaptive Boosting (AB), KNN) with the features considered above to identify either bot or a human produces the tweets. The dataset used in the experiments consists of more than 16M tweets posted by 2M different users. ET and RF achieved the best cross-validated average performance (86%), followed by DT and AB (83%) and KNN (81%).

In [4], Zi Chu et al used features extracted from the tweeting behaviour, tweet content and account proprieties to classify Twitter users into three categories: human, bot and cyborg. Authors assumed that bot's behaviour is less complex than that of humans. For that, they used an entropy rate to measure the complexity of a process, where low rates indicate a regular process; medium rates indicate a complex process, whereas high rates indicate a random process. The content of the tweet is used to build text patterns of known spam on twitter. Some other account-related features are also used in the classification such as external URLs ratio, link safety, account registration date, etc. The RF machine-learning algorithm is used to analyse these features and decide if a twitter account is a human, bot, or cyborg. The effectiveness of the classifier is evaluated through a dataset of 500,000 different Twitter users. This approach archived an average overall true positive rate of 96.0%.

## III. THE PROPOSED METHOD

### A. APPROACH OVERVIEW

The goal of the proposed approach is to classify a given Twitter account asa Bot or not, therefore, the bot accounts detection is considered as a binary classification problem with bot Twitter accounts belong to the positive class ("Bot") and normal accounts belong to the negative class ("Human"). As illustrated in Fig. 1, the bots detection system consists of several modules: the data crawling module, the graph database, the pre-processing module, and the machine-learning module. The crawling module collects Data from the Twitter Platform and stores them in a Neo4j database. The pre-processing module uses the data stored in the Neo4j database to create the training and testing dataset and the machine-learning module uses a NN classification algorithm to detectbot Twitter accounts.

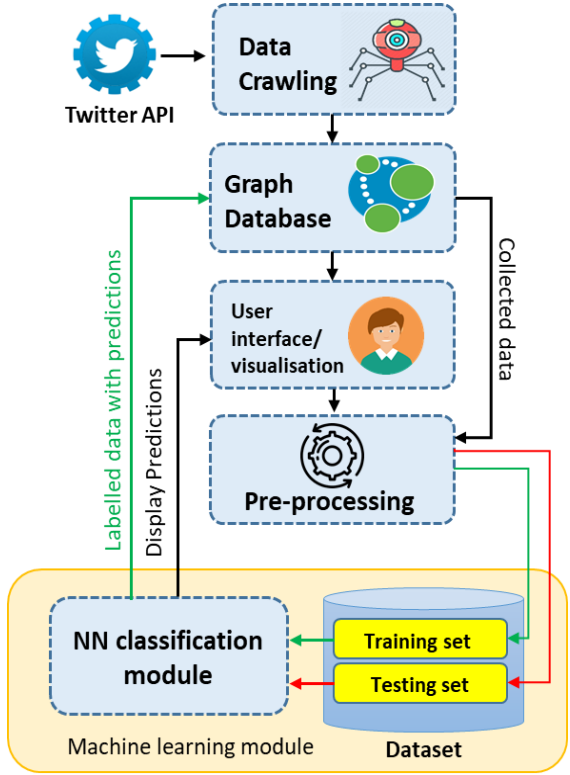

**Fig. 1**. Overview of the proposed method.

### B. DATA COLLECTION AND PROCESSING

There is no accepted worldwide dataset that can be used in the experiments, therefore, a web crawler called "PyTrawler" is developed to collect detailed user information from public accounts. PyTrawler is a Python application that uses two Python libraries, "Tweepy" to communicate with the Twitter API's and "Py2neo" to communicate with a Neo4j database and store the collected data. PyTrawler was built to handle the issue of the limited speed of data harvesting because an application can only make a limited number of requests to Twitter's API in a certain time window. As stated in the Twitter Developer Policy[1], a host is permitted 150 requests per hour. It also handles the issue of losing the internet connectivity, which gives PyTrawler the ability to continue crawling for long period of time without the need to resetting it. For each user visited, PyTrawler collected useful information about his profile, follower, friend list and posted tweets. The collected data is stored in a graph database using Neo4j technology [7]. This NoSQL database provides a browser GUI to visualise the data. The data collected and stored in the Neo4j database are formatted and normalised, then stored in a CSV file. Fig. 2 illustrates the activity diagram of PyTrawler.

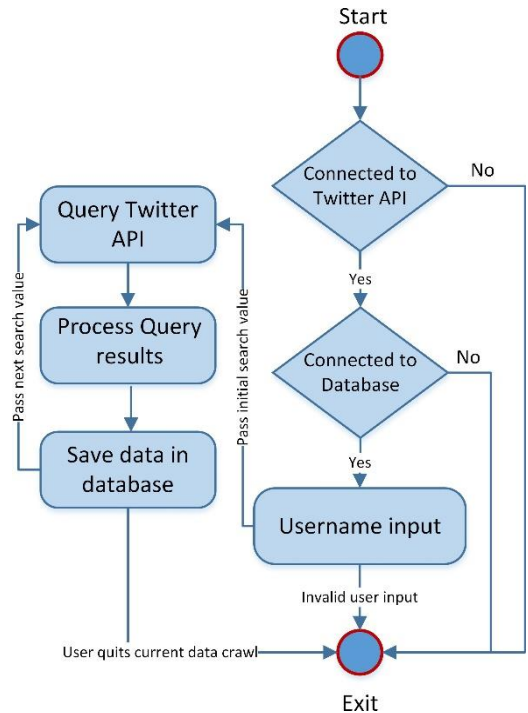

**Fig. 2.** Activity diagram of PyTrawler

A java application was developed to read raw data from the Neo4j and save the formatted data in the CSV file. The Apache Spark was used to speed up the data formatting process, where a Spark job was created to normalise all the data by using the following normalisation equation (1), which would scale each value relative to the data in the range [0, 1] for each column.

$$V(x) = Log_2 \times \left(\frac{x - min}{x + min} + 1\right) \qquad (1)$$

---
[1] https://developer.twitter.com/fr/developer-terms/agreement-and-policy

For, the training process a supervised learning approach was chosen; this involves giving the neural network a training dataset that has already been classified and labelled either 'Human' or 'Bot'. Therefore, a training and testing datasets were created, from the CSV file, to be used by the neural network. The approach used to identify bots accounts was to look at the accounts that had high TFF Ratio (Twitter Follower-Friend Ratio), then manually evaluate the account within Twitter. This is a time-consuming process but meant that the data used to train the neural network contained genuine bot accounts. The training dataset contains 760 samples labelled either 'Human' or 'bot', while the testing dataset is composed of 100 samples (humans and bots accounts).

### C. FEATURES SELECTION

Most approaches in the literature use many features. However, recent studies show that similar high performance can be achieved by using a minimal number of features [8, 2]. Thus, in our method we propose a set of nine feature that are inspired from the previous work [3, 4, 16]. The features values are gathered from the information in the NoSQL database, by examining username, account metadata, followers and friends count, description of the account, number of tweets, and content of the tweets. The explanation of the selected features is given in Table 1.

**Table 1.** Features explanation

| Feature | Explanation |
|---|---|
| Default profile | This feature has a binary value, where true, indicates that the user has not altered the theme or background of their user profile |
| Statuses count | This feature presents the number of tweets and retweets is-sued by the user |
| Follower's count | This feature presents the number of followers this account currently has |
| Listed count | This feature presents the number of Twitter lists on which this Twitter account appears |
| Friends count | This feature presents the number of users this account is following. Generally, Bot accounts have too many followings compared to human accounts |
| URLs ratio | This feature represents the frequency of included external URLs in the posted tweets by a user account. Study in [4] found that the ratio of a bot is 97%, while that of human is much lower at 29%. |
| Verified account | This feature has a binary value, whether a user account is verified by twitter or not. It is used to establish authenticity of identities of key individuals and brands on Twitter. A truevalue indicates a verified account |
| Protected account | This feature has a binary value, which indicates if the useris protecting their tweets, and they are invisible to every- one other than its selected followers, or not. Bots are mostly unprotected accounts |
| Hashtag's ratio | This feature represents the number of hashtags that a user account has used by the number of tweets. One tweet can include more than one hashtag |

### IV. EXPERIMENTAL RESULTS

In this section, we evaluate the efficiency of our classification system based on the datasets that have been previously collected. The metrics used in the evaluation are accuracy, precision, recall and F-score. The accuracy refers to the ratio of all correct predictions to the total number of input samples and it is computed by the following equation.

$$Accuracy\ (A) = \frac{TP + TN}{TP + FP + TN + FN} \quad (2)$$

Where, TP is the number of samples that are correctly classified as bot accounts, TN the number of samples that are correctly classified as human accounts and the total number of samples is the number of all samples (Bot and Human accounts). FP refers to the number of samples incorrectly classified as bot accounts and FN notes to the number of samples incorrectly classified as humans. Precision corresponds to the proportion of bot accounts that are correctly considered as bots, with respect to all correct prediction results (Bot and Human accounts) (Equation 3), while recall is the proportion of bot accounts that are correctly considered as bots, with respect to all actual bot samples (Equation 4). F-score is a weighted average between precision and recall (Equation 5).

$$P = \frac{TP}{TP + FP} \quad (3)$$

$$R = \frac{TP}{TP + FN} \quad (4)$$

$$F1 = \frac{2 \times P \times R}{P + R} \quad (5)$$

### A. EXPERIMENT SETUP

The simulation experiments were performed on a virtual machine that is running Ubuntu 18.0.4 on Intel Core i7 CPU, 3.80 GHz, with 6 GB memory. The configuration of the Multi-Layer Perceptron (MLP) neural network is built with the DL4J (Deep Learning for Java) library. DL4J is an open-source deep-learning library written in java and includes implementations of many deep learning algorithms. The layers of the MLP neural network were constructed with the *'NeuralNetConfiguration.Builder'* method from DL4J. In addition, the Apache Maven tool is used to keep the list of all required dependencies for the NN in a 'pom.xml' file, which can then be altered to import or change dependencies as needed. This meant that the setup of the project on different devices was quick and easy, which sped up the development process. Finally, a java class was developed to help the NN read the data from the CSV file produced in the pre-processing step. After the neural network was configured, it was trained with 760 samples of Twitter accounts that have already been classified and labelled either Bots or Humans. Other 100 samples are used for testing the neural net- work. The testing set represents the unknown Twitter accounts that we want to classify as Bots or Humans.

## B. TEST RESULTS

Using the training and testing datasets that had been previously created, several tests were carried out to evaluate the success of the detection method and deter-mine the accuracy of the MLP neural network. In this context, several tests were performed on the initial training of the network to find a suitable learning rate for the MLP neural network. The value of this parameter has a high influence on the training process and accuracy of the neural network, large values may result in an unstable training process, whereas small values may lead to a long training process that could get stuck.

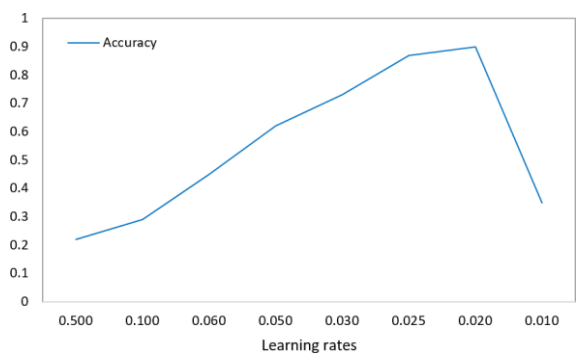

**Fig. 3.** Learning rate results on the accuracy

The results of these tests are illustrated in Fig. 3. From the obtained results, it was found that the optimum configuration was a learning rate of 0.02. From the tests results, it was found that changing the number of hidden nodes or layers did not seem to influence the accuracy of the networks predictions, so it was fixed at 25 layers. Thus, the optimum configuration used in the experiments was a learning rate of 0.02, with 200 passes and 25 hidden layers. Knowing that due to the hardware limitations on the computing power the testing was restricted to 200 passes. After the optimum initial configuration was found, the neural network was reloaded and retrained on the created training dataset. Once retrained, it was assessed against the testing dataset. The MLP neural network achieved an overall accuracy rate of 92%, which is a high rate and meets the required accuracy rate in practical use. Precision is also high with a rate of 92.59%, which shows strong overall confidence in the pattern recognition process. The recall rate was lower than the precision rate (50%), knowing that in our work, precision is more important than recall because getting False Negatives (FN), when a bot account is considered as user account, cost more than False Positives (FP), when a user account is considered as bot because a bot account could be a spam bot that was created to conduct malicious activities [3]. In this context, a study conducted by Z. Chu, et al [3] on a large Twitter dataset found that most bots include malicious URLs in tweets and that the average number of URLs in tweets sent by bots is three times higher than the number in tweets from real people accounts. The overall results of the initial test are presented in Table 2.

**Table 2.** Overall results of the initial test

| (A) | Precision (P) | Recall (R) | F-score (F1) |
|---|---|---|---|
| 92.00% | 92.59% | 50.00% | 64.94% |

## V. CONCLUSION

In this paper, we proposed a novel approach that utilises machine learning to solve the problem of bot accounts within Twitter. The proposed approach con- sists of a collection of different modules that provide functionalities to detect bots accurately and precisely within Twitter. The machine-learning module has proved that it is possible, through the analysis of account data, to detect bot accounts within Twitter by only looking at a small number of account features. Despite the neural network not being 100% accurate, it was able to learn and improve itself with minor supervision.

In the future, we plan to improve this work by the use of more samples to properly train and test the neural network, which will with no doubt enhance the predictive accuracy of the classifier. Furthermore, we intend to implement multiple neural networks that looked at different aspects of the data that were collected from Twitter and compare the results from the applied deep learning algorithms.

ACKNOWLEDGEMENT.

This project has received funding from the European Union's Horizon 2020 research and innovation programme under grant agreement no. 786698. This work reflects authors' view and Agency is not responsible for any use that may be made of the information it contains.